\documentclass[showpacs,amssymb,amsmath,nobibnotes, aps,prl,secnumarabic
,twocolumn%
]{revtex4}
\pdfoutput=1
\usepackage{bm}
\usepackage{natbib}
\usepackage{graphicx}

\begin{document}
\textfloatsep 10pt

\title{Parametrically excited water surface ripples as ensembles of oscillons}

\author{M. Shats}
\email{Michael.Shats@anu.edu.au}
\author{H. Xia}
\author{H. Punzmann}

\affiliation{Research School of
Physics and Engineering, The Australian National
University, Canberra ACT 0200, Australia}

\date{\today}

\begin{abstract}
We show that ripples on the surface of deep water which are driven parametrically by monochromatic vertical vibration represent ensembles of oscillating solitons, or quasi-particles, rather than waves. Horizontal mobility of oscillons determines the broadening of spectral lines and transitions from chaos to regular patterns. It is found that microscopic additions of proteins to water dramatically affect the oscillon mobility and drive transitions from chaos to order. The shape of the oscillons in physical space determines the shape of the frequency spectra of the surface ripple.

\end{abstract}

\pacs{47.35.Pq, 47.27Cn, 47.52.+j}

\maketitle

Ripples on the surface of vertically vibrated liquids, often referred to as Faraday waves to honor his pioneering effort \cite{Faraday1831}, are commonly used to study generation of regular patterns (e.g. \cite{Cross1993, Kudrolli1996}) and wave turbulence (e.g. \cite{Wright1996, Snouck2009}). Surface ripples are usually explained by nonlinear interactions of waves in the gravity-capillary range \cite{Miles&Henderson1990, Perlin_Schultz2000}.

Recent experimental studies of parametric surface waves revealed several problems with the wave interpretation of the disordered surface ripples. Simultaneous measurements of the frequency and wave number spectra \cite{Snouck2009} reveal poor correspondence between the two: frequency spectra show distinct multiple harmonics, while the wave number spectra are continuous. It has also been found that the dispersion relation is satisfied only approximately. The origin of the universally observed multiple frequency harmonics and the mechanism of their spectral broadening also remain poorly understood. It has been shown that in the time domain, initially continuous waves appear to be broken into ensembles of "envelope solitons" by modulation instability \cite{Punzmann2009, Shats2010, Xia2010}. What remains unclear is how these solitons are realized in physical space.

In this Letter we show that the surface ripples in a parametrically driven system consist of oscillating solitons, or oscillons, previously found in vertically vibrated flows \cite{Wu_1984,Umbanhowar1986,Lioubashevski_1996, Lioubashevski_1999, Arbell1998}. We also show that the seemingly chaotic state on the water surface can be turned into an ideally ordered matrix of oscillons by reducing their mobility.

\begin{figure}[b]
\includegraphics[width=6.0 cm]{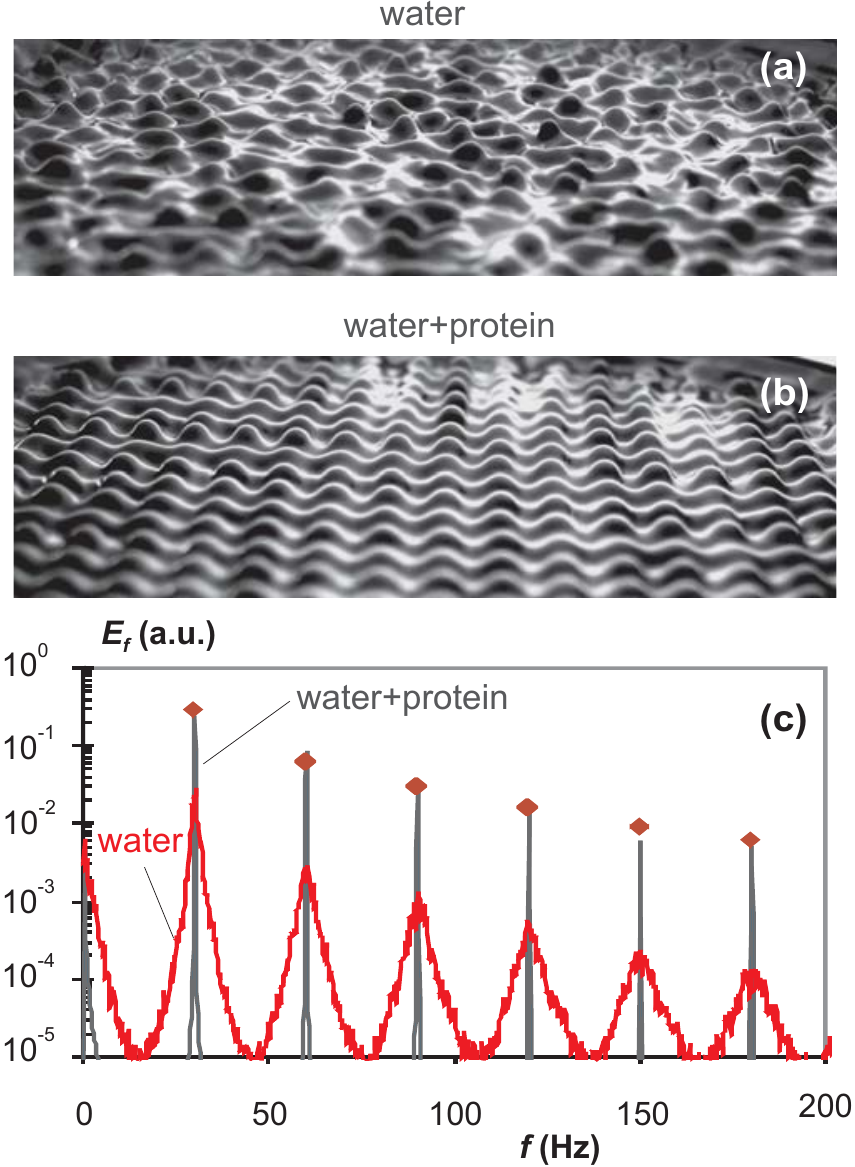}
\caption{\label{fig1} (a,b) Photographs of the surface ripples under parametric excitations at 60 Hz and acceleration of $a=1.2g$ in (a) water, and (b) water with a small (0.5 p.p.m. in weight) addition of the bovine serum albumin. (c) Frequency spectra of the surface gradient measured using reflection of a thin laser beam off the fluid surface in water (red) and with albumin added (black).}
\end{figure}

We report results obtained in a vertically shaken circular container (178 mm diameter, 30 mm deep) with monochromatic forcing. The experimental setup and the measurement techniques are described in references \cite{Punzmann2009, Shats2010, Xia2010}. Waves are generated in the range of frequencies of $f_0$ = (10-200) Hz. Figure~\ref{fig1}(a) shows a photograph of the surface ripple in water, forced at $f_0$ = 60 Hz at a peak-to-peak acceleration of $a$ = 1.2$g$. The wave field looks random; it is dominated by oscillating "blobs" approximately 4 mm in diameter. Blobs move randomly on the surface; they merge and collide with other blobs. The frequency spectrum of the surface gradient, measured using reflection of a very thin (0.5 mm) laser beam off the surface, is not random. It consists of spectrally broadened harmonics, $f_n=nf_0/2$, Fig.~\ref{fig1}(c).

For the same vertical acceleration of $a = 1.2g$, the surface ripple changes dramatically when a very small quantity of a protein suspension (skim milk, gelatine and albumin were tested) is added to water, as shown in Fig.~\ref{fig1}(b). In this example, we have added 0.5 p.p.m. (in weight) of bovine serum albumin to water. Such microscopic additions of proteins do not change surface tension ($\sigma = 73$ mN/m) and viscosity ($\eta = 1 mPa\cdot s$ at 20C) of water. However, the resulting wave field resembles standing waves which form very stable square pattern. The frequency spectrum does not change qualitatively, but spectral harmonics become very narrow, as seen in Fig.1(c). Red diamonds in Fig.~\ref{fig1}(c) show the spectral powers contained in each spectrally broadened harmonic in water (integral over $\Delta f = f_n \pm 15$ Hz). The spectral energy contained in each harmonic in water does not change with the addition of proteins compared with the peaks in spectra in water/protein solution. This is consistent with the fact that forcing (acceleration) and dissipation (viscosity) remain constant.

\begin{figure}
\includegraphics[width=6.0 cm]{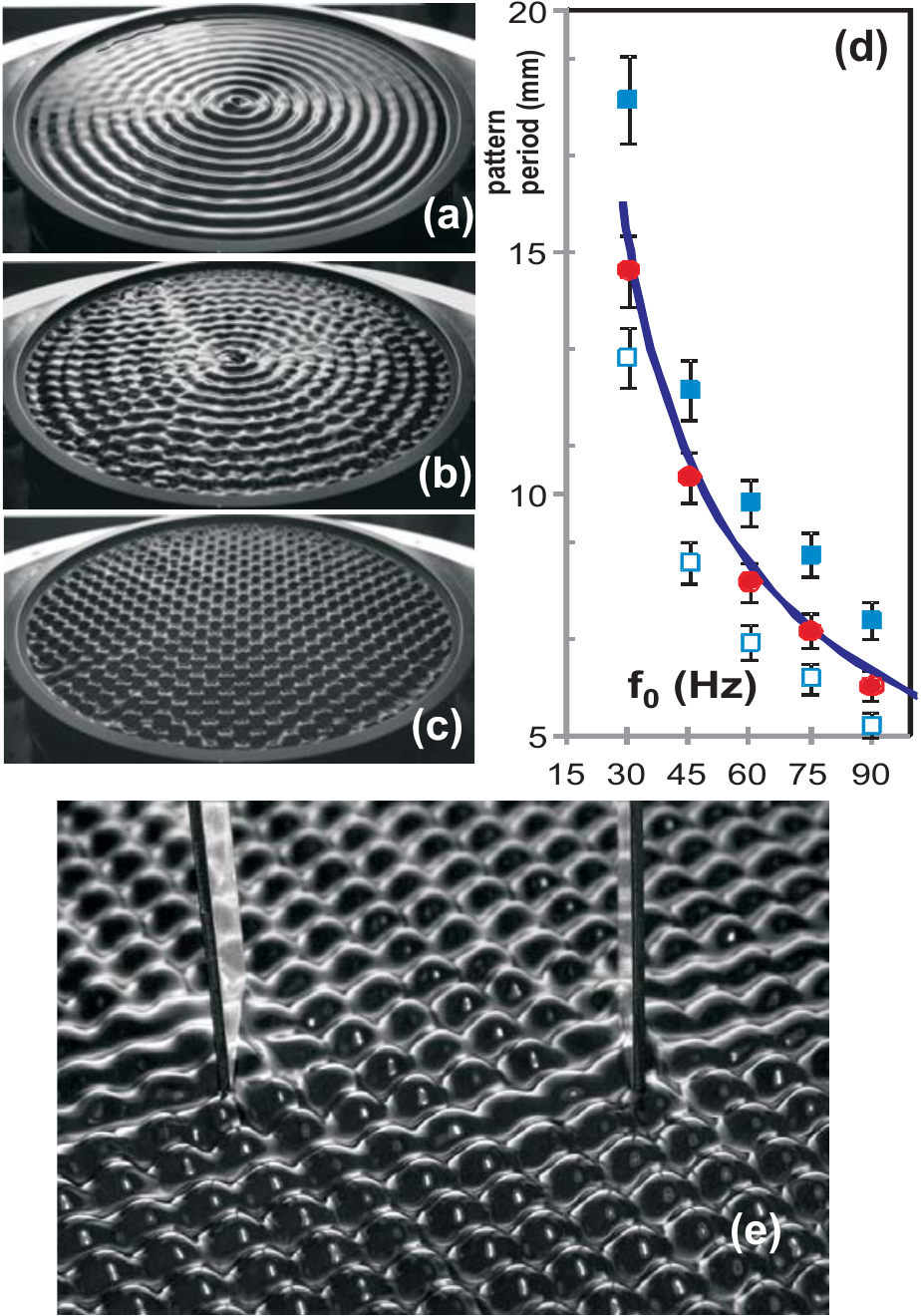}
\caption{\label{fig2} Flow development during the startup phase in the albumin solution at (a) $t = 1.5$ s, (b) $t = 2.5$ s, and (c) $t = 6.0$ s. (d) Spatial period of a square pattern as a function of driving frequency: solid squares correspond to the distance between the maxima along the diagonal of the matrix; open squares correspond to the distances in rows/columns. Red circles show spatial periods of the concentric circles at the early stage of the ripple evolution. Solid line shows dispersion relation of the capillary-gravity waves. (e) A square pattern is not substantially affected by the insertion of metal plates into the established matrix.}
\end{figure}

The dynamics of the formation of the square pattern in the albumin water solution is illustrated in Figure~\ref{fig2}. Shortly after the shaker is turned on, a standing wave in the round container forms circular wave fronts, Fig.~\ref{fig2}(a). Later these concentric circles become modulated azimuthally, Fig.~\ref{fig2}(b). The standing wave appears to be broken into an ensemble of blobs, which eventually self-organize into a square matrix in steady state, Fig.~\ref{fig2}(c). To test if the matrix periodicity is consistent with the capillarity-gravity wave dispersion relation, we perform measurements at different forcing frequencies. The distances between maxima in the circular standing waves agree well with the dispersion, confirming their wave nature, Fig.~\ref{fig2}(d). However, the distances between adjacent peaks in a square matrix noticeably deviate from the wave lengths expected from linear dispersion relation $\omega^2=gk+(\sigma/\rho)k^3$, where $k$ is the wave number and $\rho$ is the density. The discrepancy between the blob periodicity and the dispersion relation suggests that the observed pattern is not a standing wave. It is a matrix of non-propagating spatially localized oscillating surface perturbations. Indeed, the insertion of two metal plates into the liquid does not destroy a global pattern [Fig.~\ref{fig2}(e)], as should happen with standing waves sensitive to the boundary conditions.

Oscillating blobs, which organize into a square pattern, resemble oscillating solitons, or oscillons. Such localized oscillatory perturbations of the liquid surface have been studied since 1984. The first parametrically driven stationary solitonic structures were discovered on the water surface in a resonator \cite{Wu_1984}. Later oscillons, were found in granular layers \cite{Umbanhowar1986}, in thin layers of highly dissipative fluids \cite{Lioubashevski_1996}, in non-Newtonian fluids \cite{Lioubashevski_1999}, and in strongly dissipative liquids vibrated at two frequencies \cite{Arbell1998}. Recently a non-propagating solitary wave on the water surface was observed in a very narrow vertically vibrated cell below the threshold of parametric excitation by externally perturbing the surface \cite{Rajchenbach2011}. The dissipation in that system was high due to strong wall friction.

\begin{figure}
\includegraphics[width=6.0 cm]{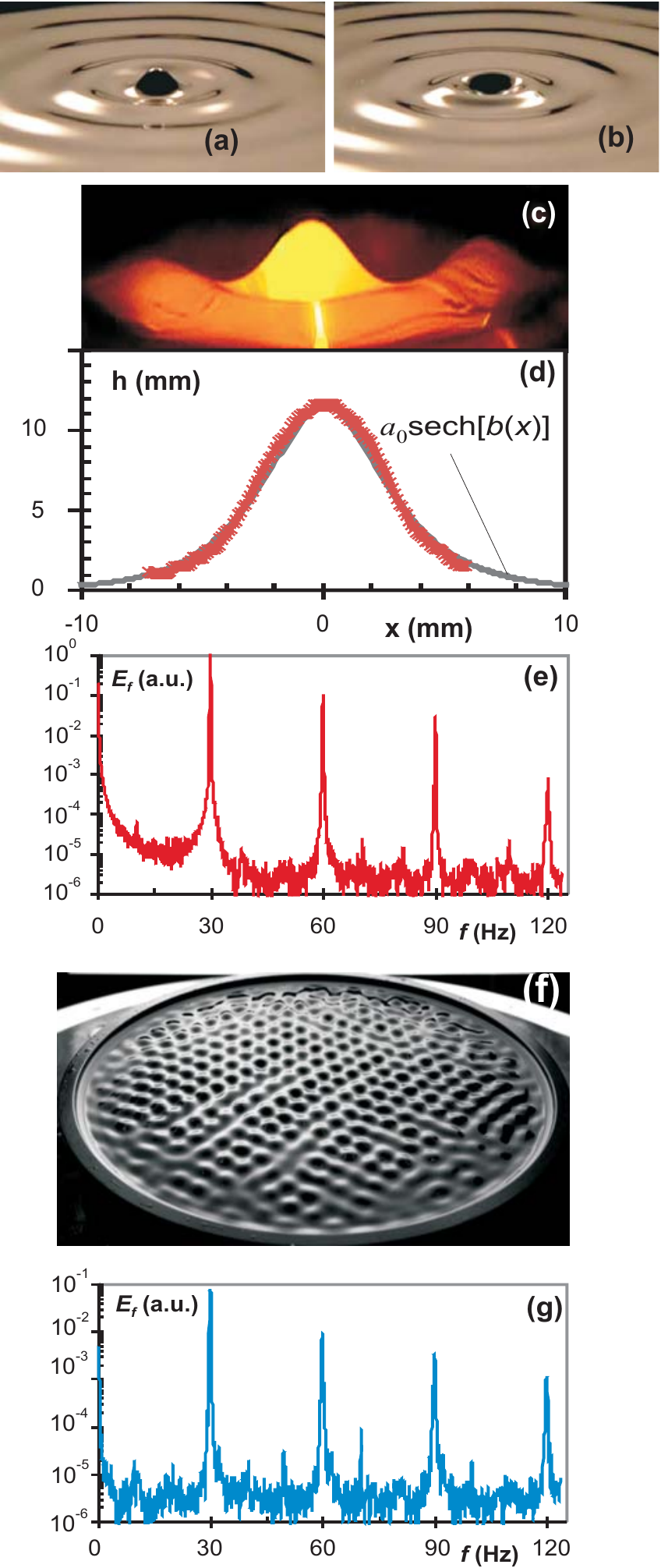}
\caption{\label{fig3} (a,b) A single oscillon excited at the subcritical acceleration ($a=3.1g$) in viscous water solution of glycerine by external perturbation at 30 Hz. (c) A snapshot of the oscillon profile illuminated by a vertical laser sheet after adding fluorescent dye to the glycerine-water solution. (d) surface elevation of the oscillon is well approximated by the hyperbolic secant fit. (e) Frequency spectrum of the surface gradient of a single subcritical oscillon. (f) Surface ripple on the glycerine-water solution above critical excitation ($a=3.3g$). (g) Frequency spectrum of the surface gradient of the supercritical ripple. The excitation frequency is 60 Hz.}
\end{figure}

Here we report the first observation of an oscillon in a 3D flow driven monochromatically on the surface of a deep layer of liquid (larger that the wave length). First, we study a single isolated oscillon in a highly viscous water solution of glycerine ($\eta = 190 mPa\cdot s$). A shaker is driven at the frequency of $f_0 = 30$ Hz at the acceleration just below the threshold of parametric excitation, $a = 3.1g$. The liquid surface stays flat until it is perturbed externally (e.g. by a falling droplet), as seen in the frame sequence of Fig.~\ref{fig3}(a,b) obtained from the fast video (300 fps) movie \cite{movie_oscillon}. A localised oscillating structure is sustained by the forcing for a very long time (over several minutes).

The shape of the oscillon was studied by illuminating its cross section by a thin green laser sheet in combination with a fluorescent dye (Rhodamine B) which was added to the liquid. The laser sheet produces a sharp image, as shown in Fig.~\ref{fig3}(c). The profile of the surface contour is well approximated by the hyperbolic secant fit, Fig.~\ref{fig3}(d). This is in agreement with the oscillon shape in other systems (e.g. \cite{Wu_1984}).
The frequency spectrum of the surface gradient of a single subcritical oscillon is shown in Fig.~\ref{fig3}(e) for the excitation frequency of $f_0 = 60$ Hz. Remarkably, this spectrum shows the subharmonic of the excitation frequency as well as its harmonics. Multiple harmonics are observed universally in all experiments with parametrically excited waves. Their presence is usually attributed to wave-wave interactions or even to the energy cascade in wave turbulence (e.g. \cite{Snouck2009}). Here we show that multiple harmonics is a feature of an individual oscillon.

\begin{figure}
\includegraphics[width=6.0 cm]{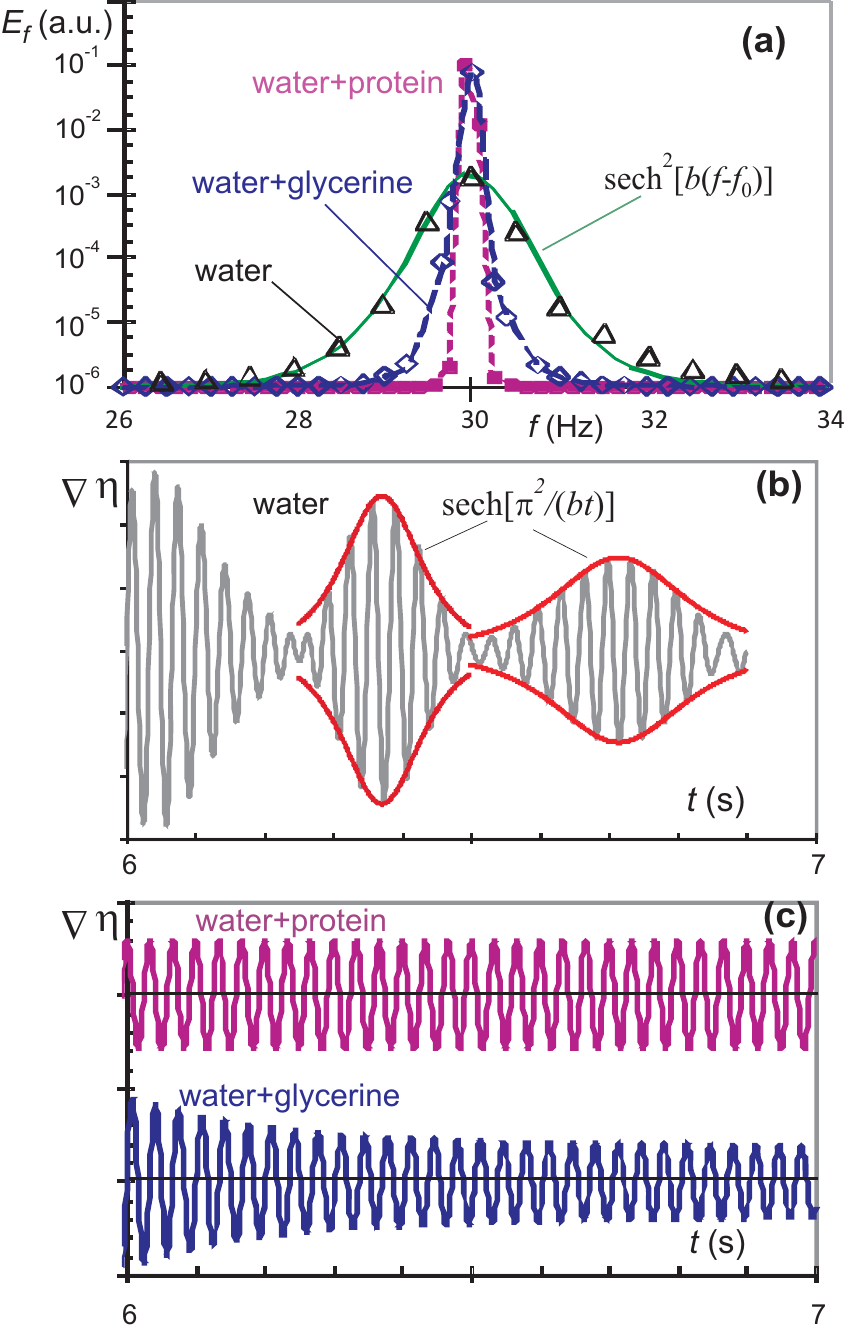}
\caption{\label{fig4} Shapes of the spectral lines on the water surface illustrated in Fig.~\ref{fig1}. (a) Frequency spectral line of the first subharmonic (30 Hz) in water (triangles), water-glycerine (diamonds), and in water-albumin solutions (squares). Solid line shows $sech^2$ fit. Waveforms of the surface gradient at the fixed point on the surfaces of (b) water, and (c) water-glycerine and water-albumin solutions.}
\end{figure}

When the acceleration is increased above the threshold of the parametric excitation, multiple oscillons are generated. They form a square matrix, Fig.~\ref{fig3}(f), identical to those observed in water with added protein (Fig.~\ref{fig2}). Their frequency spectra measured on the surface of the oscillon matrix, Fig.~\ref{fig3}(g), are very similar to those of individual oscillons, Fig.~\ref{fig4}(c). Thus, frequency spectra of the surface motion are determined by individual oscillons and not by collective effects.

At higher forcing, oscillons become more mobile in horizontal plane. This increased mobility gradually leads to a less regular pattern, as seen in a fast video \cite{movie_matrix}. A difference in the mobility of oscillons determines the spectral widths of the frequency harmonics. Since the shape of an oscillon in physical space is given by the hyperbolic secant,  $h(x) \sim sech(ax)$, its slow (compared with the period of oscillation) movement about the observation point should lead to a modulation envelope $h(t) \sim sech[\pi/(at)]$ in the time domain. This, in turn, should lead to a spectral line shape in a frequency spectrum given by the squared hyperbolic secant function of frequency, $E(f) \sim sech^2[b(f-f_0)]$ \cite{Punzmann2009}. The analysis of the surface ripples of Figs.~\ref{fig1} and \ref{fig3} confirms these expectations. Fig.~\ref{fig4}(a) shows shapes of the spectral lines (first subharmonic) in pure water and in the albumin solution (same conditions as in Fig.~\ref{fig1}). A higher mobility of oscillons in water leads to the line shape $E(f) \sim sech^2[b(f-f_0)]$. The addition of albumin, reduces the line width to the instrumental limit determined by the measurement time. In the time domain, the higher oscillon mobility in water leads to the observation of the "envelope solitons" previously reported in \cite{Punzmann2009, Xia2010}. The results presented here reveal that the hyperbolic secant envelopes result from the shape of the individual oscillons in physical space. The $sech$ analytic fit to the time-domain signal is illustrated in Fig.~\ref{fig4}(b). In contrast to this and consistent with the narrow spectrum, stationary solitons within the matrix [Figs.~\ref{fig1}(b), ~\ref{fig2}(c), ~\ref{fig3}(f)] generate non-modulated harmonic signals seen in Fig.~\ref{fig4}(c). The results similar to those obtained in the water-protein suspension have been found in a viscous glycerine-water solution, as illustrated in Fig.~\ref{fig4}. This suggests that the same effect, namely the reduction of the oscillon mobility (possibly due to the change in their short-range interaction) is achieved in Newtonian (glycerine) and possibly non-Newtonian (water-protein) liquids.

Oscillons result from the instability of the surface waves, as illustrated in Figs.~\ref{fig2}(a-c). This instability in the capillary wave range is modulation instability \cite{Xia2010}, a counterpart of the well known Benjamin-Feir instability of gravity waves \cite{Benjamin1967}. In its nonlinear stage, this instability leads to the breaking of waves into ensembles of oscillons, Fig.~\ref{fig2}. A similar process is observed in the gravity wave range in large wave tanks, where the cross wave instability destroys planar wave fronts \cite{Lichter1986}. The connection between cross waves and the oscillon formation has been suggested earlier \cite{Larraza1984}. Since oscillons are formed in the process of breaking of the surface wave, they inherit their size from the mother wave such that their diameter and the period of patterns that they form roughly follow the dispersion relation, Fig.~\ref{fig2}(d).

Individual, or subcritical, oscillons shown in Fig.~\ref{fig3} were observed in a viscous glycerine water solution. They have also been observed in water with added protein suspensions, though their amplitudes were lower there. Stronger external perturbations in water lead to the onset of multiple oscillons on the surface and to a pattern formation.  A mechanism through which microscopic additions of proteins (typically less than 1 p.p.m.) dramatically change the mobility of oscillons is not understood. It is well known, however, that small additions of high molecular weight polymers have striking effects on the stability and pattern selection in turbulence, for example in Couette-Taylor flow \cite{Groisman1996}. This is attributed to different viscoelastic properties of these liquids. As we show here, oscillons are observed in both Newtonian (glycerine) and in weakly non-Newtonian fluids. The effect of proteins on the patterns of surface ripples should be further investigated.

Our results suggest that patterns and ripples are made of oscillons which interact on a short range.
There are two types of subharmonic oscillons which differ by a phase shift of $1/f_0$. As in other oscillon systems (e.g. \cite{Umbanhowar1986}), the like-phase oscillons show repulsive interaction, while oscillons of the opposite phase attract and bind. Such interaction forces lead to a variety of effects, including the formation of nontrivial patterns, e.g. square matrices in a circular container, Figs.~\ref{fig2}(c),~\ref{fig3}(f), as well as stripes (when the matrix symmetry breaks along rows or columns).

It should be noted that the secant hyperbolic solitons are known as a limiting case of cnoidal waves. A transition from a periodic cnoidal wave to a \textit{sech} solitary structure determined by the elliptical parameter $\kappa$ ($\kappa \lesssim 1$ for waves and $\kappa =1$ for solitons) has been discussed in relation to the oscillon theory \cite{Miles1984}.

The interpretation of the spectral broadening as the result of the increased mobility of oscillons may be relevant in other wave systems as well. Exponential frequency spectra are observed universally in many parametrically excited systems, such as the magnetic spin waves excited in ferrites \cite{Krutsenko1978}, nonlinear waves in optical fibres \cite{Kutz2005}, Raman fibre lasers \cite{Babin2008} and in magnetically confined plasma \cite{Pace2008}. Modeling of the parametrically driven damped nonlinear Schr\"{o}dinger equation \cite{Wang2001} shows the development of nonlinear localized structures, such as oscillons in dissipative nonlinear systems.

Finally, in his seminal paper \cite{Faraday1831}, M. Faraday mostly refers to "crispations" made of "heaps" rather than waves.

\begin{acknowledgments}
This work was supported by the Australian Research Council's Discovery Projects funding scheme (DP110101525).
\end{acknowledgments}

\end{document}